\documentclass[a4paper]{jpconf}
\usepackage{graphicx}
\usepackage{amsmath}
 \usepackage{amscd}
 \usepackage{hyperref}
 \usepackage{amssymb}
\usepackage{epsf,float}
\usepackage{bm}%

\def\beq{\begin{equation}}
\def\eeq{\end{equation}}
\def\beqa{\begin{eqnarray}}
\def\eeqa{\end{eqnarray}}
\def\beg{\begin{lyxgreyedout}}
\def\eeg{\end{lyxgreyedout}}

\begin{document}
 \title{On the intrinsically cyclic nature of space-time in elementary particles}

\author{Donatello Dolce}
\address{The University of Melbourne, School of Physics (CoEPP), Parkville VIC 3010, Australia.}
\ead{ddolce@unimelb.edu.au}
\begin{abstract}
We interpret the relativistic quantum behavior of elementary particles in terms of elementary cycles. This represents a generalization of the de Broglie hypothesis of intrinsically ``periodic phenomenon'', also known as ``de Broglie internal clock''. Similarly to a ``particle in a box'' or to a ``vibrating string'',  the constraint of intrinsic periodicity represents a   semi-classical quantization condition, with remarkable formal correspondence to ordinary relativistic quantum mechanics. In this formalism  the retarded local variations of four-momentum characterizing relativistic interactions can be equivalently expressed in terms of retarded local modulations of de Broglie space-time periodicity, revealing a geometrodynamical  nature of gauge interaction.
\end{abstract}

\section{Introduction}

 In this paper we discuss how relativistic bosonic fields, similarly to the harmonics of a ``vibrating string'' or to the semi-classical quantization of a ``particle in a box'',  can be consistently quantized by imposing their characteristic de Broglie space-time  periodicity as dynamical constraint \cite{Dolce:2009ce,Dolce:tune,Dolce:AdSCFT}, see also \cite{Dolce:2009cev4,Dolce:Dice,Dolce:2010ij,Dolce:2010zz,Dolce:FQXi}. 
Such an assumption of intrinsic periodicity can be regarded as a natural realization of the de Broglie hypothesis at the base of undulatory mechanics (wave-particle duality) \cite{Broglie:1924,Broglie:1925}. 
By quoting the introduction of de Broglie's famous PhD thesis, the formalism described in this paper is based on the fundamental assumption ``\textit{of existence of a certain periodic phenomenon of a yet to be determined character, which is to be attributed to each and every isolated energy parcel [elementary particle]}'' \cite{Broglie:1925,Broglie:1924}.  This so-called ``de Broglie periodic phenomenon''  \cite{1996FoPhL}  has been implicitly tested by 80 years of QFT and indirectly observed in a recent experiment \cite{2008FoPh...38..659C}. 
A similar assumption was also used, for instance, by Schr\"odinger in his formulation of the \emph{Zitterbewegung} model of the electron,  by Bohr in his description of the Hydrogen atom,  by Sommerfeld for his quantization condition, etc. 
Moreover, as Galileo taught us with his study on the isochronism of the pendulum, a consistent formalization  of the concept of time in physics requires  an assumption of intrinsic periodicity for isolated elementary systems. Time can only be defined by counting
the number of cycles of phenomena  \textit{supposed} to be
periodic (this guarantees the invariance of the unit of time). Indeed the operative definition of a second is ``the duration of 9,192,631,770 characteristic cycles
of the Cs atom'', where $T_{Cs}\sim10^{-10}s$.  Thus, according to  de Broglie's assumption of  ``periodic phenomenon'' and considering Newton's law of inertia, \emph{every elementary particle} constituting a physical system (atomistic description) \emph{can be regarded as a reference clock}, the so-called  ``de Broglie internal clock'' \cite{2008FoPh...38..659C}.  
 
To illustrate the idea we consider in this introduction only  periodicity in time $T_t$. According to the de Broglie assumption of ``periodic phenomenon'',  to an elementary ``parcel'' of energy $\bar{E}(\mathbf{\bar{p}})$, observed in a generic reference frame denoted by $\mathbf{\bar{p}}$, there  is associated 
a de Broglie time  periodicity $T_{t}(\mathbf{\bar{p}})=h/\bar{E}(\mathbf{\bar{p}})$, 
\cite{Broglie:1924,1996FoPhL,2008FoPh...38..659C}.
 Relativistic causality is preserved by the assumption of intrinsic periodicity because every retarded and local variation of energy $\bar{E}(\mathbf{\bar{p}})$ can be  equivalently described  in terms of retarded and local variation of the dynamical de Broglie periodicity $T_{t}(\mathbf{\bar{p}})$ of the particle.
The definition of relativistic clock given by A. Einstein \cite{Einstein:1910} is: ``\textit{by a
clock we understand anything characterized by a phenomenon passing
periodically through identical phases so that we must assume, by the
principle of sufficient reason, that all that happens in a given period
is identical with all that happens in an arbitrary period'}'. This means that the
whole information  of a relativistic clock, and thus of the de Broglie ``internal clock'' of a particle, is contained in a single
 (space-time) period.   
Therefore, by using the language of extra dimensional theories \cite{Dolce:AdSCFT}, we formalize the intrinsic de Broglie ``periodic phenomena'' in terms of fields embedded  in a compact time dimension  of relativistic length
$T_{t}(\mathbf{\bar{p}})$ and Periodic Boundary Conditions (PBCs) \cite{Dolce:2009ce}. For the sake of simplicity we consider only the bosonic case. 
Thus the intrinsically periodic field solution of our bosonic theory  is actually similar to a vibrating string embedded in a cyclic time dimension or to a particle in a box. Through discrete Fourier transform,
to a compact variable corresponds a quantized conjugate variable,
\emph{i.e.} a variable which takes discrete values.  Considering the 
relation $\bar{E}(\mathbf{\bar{p}})=\hbar\bar{\omega}(\mathbf{\bar{p}})$, to the specific case of an elementary  isolated
system with intrinsic periodicity $T_{t}(\mathbf{\bar{p}})$  is associated  the quantized energy spectrum
$E_{n}(\mathbf{\bar{p})}=n\hbar\bar{\omega}(\mathbf{\bar{p}})=nh/T_{t}(\mathbf{\bar{p}})$.  PBCs imply that the only possible
energy eigenmodes are those with an integer number of cycles, so that we have a correspondence with
the Bohr-Sommerfeld quantization condition (for instance, it can be shown that the periodicity
condition $E_{n}T_{t}=nh$ can be more in general written as $\oint E_{n}dt=nh$ for interacting systems).
This allows one to solve non-relativistic quantum problems in a semi-classical way \cite{Dolce:2009ce,Dolce:2009cev4}.

 Similarly to the relativistic Doppler effect we must consider that  a de Broglie ``periodic phenomenon'' appears to have different periodicities if observed from  different reference frames denoted by spatial momentum $\mathbf{\bar{p}}$. The periodicity of a ``periodic phenomenon'' varies with  the energy $\bar{E}(\mathbf{\bar{p}})$ associated to the corresponding particle.  
According to $T_{t}(\mathbf{\bar{p}})=h/\bar{E}(\mathbf{\bar{p}})$, the de Broglie time periodicity
must be regarded as local and dynamical as the energy. Therefore, it varies according to relativistic causality. Since the time periodicity
$T_{t}(\mathbf{\bar{p}})$ transforms in a relativistic way, the proper-time intrinsic
periodicity $T_{\tau}=T_t(0)$ fixes the upper bond of the time periodicity  $T_{\tau}\geq T_{t}(\mathbf{\bar{p}})$. In fact the mass $\bar M = h / T_\tau c^2$ is the lower bond of the energy, $\bar{M}c^{2}\leq\bar{E}(\mathbf{\bar{p}})$.  The proper time periodicity $T_{\tau}$ is  the time for light to travel across the Compton wavelength $\lambda_s  = T_\tau c $ of the particle. 
The heavier the
mass, the faster the proper-time periodicity of the de Broglie ``periodic phenomenon''. 
Hence, even a light particle
such as the electron has  intrinsic
time periodicity equal or faster than $\sim10^{-20}s$.
It is important to note that this periodicity
is about ten orders of magnitude away from the characteristic time periodicity
of the cesium atomic clock
$T_{Cs} \sim 10^{-10}s$ (the difference between these two periodicities is of the order of the difference between a solar year and the age of the universe). This is extremely fast even if compared with the present
experimental resolution in time ($\sim10^{-17}s$).  
Thus, for every known  matter particle (except neutrinos) we are in the case of too
fast periodic dynamics. This aspect provides a link to one of the main motivations for our assumption of a cyclic nature
of space-time in elementary particles, \emph{i.e.}  the 't Hooft determinism  \cite{'tHooft:2001ar,Elze:2002eg}. It states that there is a ``\textit{close relationship between the quantum harmonic
oscillator}'' with angular frequency $\bar{\omega}=2\pi/T_{t}$, \emph{i.e.}
 the single mode of an ordinary
quantum field with energy $\bar{E}=\hbar\bar{\omega}$, ``\textit{and a classical particle moving  along a circle of periodicity
$T_{t}$}''. In fact, if the periodicity is too fast with respect to our resolution in time, it turns out that
at every observation the system appears in an aleatoric phase of its
cyclic evolution. This is like observing a
``clock under a stroboscopic light'' \cite{Elze:2002eg}. The idea of these deterministic models is that, due to the extremely fast cyclic dynamics,
we loose information about the underlying classical theory and we
observe a statistical theory --- thus described by fields --- that matches QM. For this reason we speak
about deterministic or pre-quantum theories. The ``de Broglie
intrinsic clock'' of elementary particles can also be imagined as a
``\textit{de Broglie deterministic dice}'' \cite{Dolce:Dice}; that is a dice rolling with extremely fast de Broglie
time periodicity $T_{t}$ with respect to our resolution in time, so that the outcomes can only be described statistically. Similarly to  't Hooft's determinism we will see that the statistic description of such a fast cyclic behavior has remarkable correspondences with ordinary QM.

We may also note that, on a cyclic geometry such as that of the de Broglie ``periodic phenomenon'', there are many possible classical paths characterized by different winding  numbers
between every initial and final point. That is, a field with PBCs can self-interfere.  
Its evolution is naturally described by a sum over the  classical paths associated to its cyclic behavior. \emph{This gives rise to a remarkable formal correspondence to the ordinary
Feynman Path Integral} (FPI) of QM. Moreover, because of the cyclic nature of the ''periodic phenomenon'', \emph{the theory has implicit commutation relations}.
  The reader interested in more details or in the mathematic proofs may refers to \cite{Dolce:2009ce,Dolce:2009cev4}. In this paper we will also introduce some new aspects of the theory recently published \cite{Dolce:tune}, \emph{i.e.} the generalization of the results of \cite{Dolce:2009ce} to gauge interaction.  The idea is that every local and retarded variation of four momentum associated to a relativistic interaction scheme can be equivalently described as local and retarded modulation of the space-time periodicity of a corresponding de Broglie "intrinsic clock". In turn, the variations of periodicity can be encoded in corresponding deformations of the space-time of the theory. The geometrodynamical description of interaction arising from this formalism has interesting correspondences with general relativity. In fact, linearized gravitational interaction can be equivalently described as modulations of (space-time) periodicity of reference clocks (time dilatation and length contraction), which in turn are encoded in corresponding deformations of the metric \cite{Ohanian:1995uu}.

\section{Compact Space-Time formalism}

In classical-relativistic mechanics every isolated elementary system is described by a persistent four-momentum $\bar{p}_{\mu}=\{\bar{E}/c,- \mathbf{\bar{p}}\}$ (Newton's principle of inertia). On the other hand, 
 undulatory mechanics prescribes that a four-momentum
can be equivalently described in terms of de Broglie four-angular-frequency, according to the relation $\hbar \bar{\omega}_{\mu}  =\bar{p}_{\mu}c$.
Here we will apply (literaly) the de Broglie assumption of ``periodic phenomenon'' by describing elementary systems of (classical) four-momentum $\bar p_{\mu}$ in terms of intrinsically ``periodic phenomenon'' whose
periodicity is the usual de Broglie space-time periodicity $T^{\mu}=\{T_{t},\vec{\lambda}_{x}/c\}$. As noticed by de Broglie, a  ``periodic phenomenon'' with mass $\bar M$ is fully characterized by the corresponding proper-time periodicity $T_{\tau}= h / \bar M c^2$ \cite{Broglie:1925,Broglie:1924}.  This means that it has topology $\mathbb{S}^{1}$.  In fact, in a generic frame, the spatial and temporal components of the de Broglie four-periodicity $T^{\mu}$  are obtained through Lorentz transformations from the proper-time periodicity: $cT_{\tau} = c \gamma T_{t} - \gamma \vec \beta \cdot \vec \lambda_{x}$. The energy and  momentum of a particle with mass $\bar M$ in the new reference frame is $E= \gamma \bar M c^{2}$ and $\mathbf{\bar{p}} = \gamma \vec \beta \bar M c$, respectively. 
Therefore, in a generic reference frame,  we have de Broglie-Planck relation (de Broglie phase harmony)
\begin{equation}
c^{2} \bar M  T_{\tau} \equiv h ~~~~~~~~~\rightarrow ~~~~~~~~~ c \bar p_{\mu} T^{\mu}\equiv h \,.\label{eq:PdB:relation}
\end{equation}

As the Newton's law of inertia doesn't imply that every point particle
moves on a straight line, our assumption of intrinsic periodicities
does not mean that a system of elementary particles  should appear to be periodic.
The local, retarded modulations of four-momentum occurring during interaction and characterizing relativistic causality can be equivalently described as of local, retarded modulations of the de Broglie space-time periodicity.
This dynamical relation between space-time periodicity and energy-momentum guarantees time ordering and relativistic causality in the theory.  A combination of ``periodic phenomena'' in general forms an ergodic system. Furthermore, if also interaction is considered, the system results to have very chaotic evolutions. This yields interesting considerations about the arrow of time \cite{Dolce:2009ce,Dolce:FQXi}.   

The free cyclic field $\Phi(\mathbf{x},t)$
can be written as a tower of energy eigenmodes $\phi_n(x)$ 
\begin{equation}
\Phi(\mathbf{x},t)=\sum_{n}\mathcal{N}_{n}\alpha_{n}(\bar{\mathbf{p}}) \phi_{n}(\mathbf{x})u_{n}(t)~,~~~~~~~~\text{where:}~~~u_{n}(t)=e^{-\frac{i}{\hbar} E_{n}(\mathbf{\bar{p}})t}\,, ~~E_{n}(\bar{\mathbf{p}})=n\hbar\bar{\omega}(\mathbf{\bar{p}}).\label{field:exp:modes}\end{equation}
In this simple description  $\mathcal{N}_{n}$  is the normalization factor and   $\alpha_{n}(\bar{\mathbf{p}})$ is the  coefficients of the Fourier expansion. 
By bearing in mind the relation $\bar{E}(\bar{\mathbf{p}})=\hbar\bar{\omega}(\mathbf{\bar{p}})$,
the quantized energy spectrum $E_{n}(\bar{\mathbf{p}}) = n \bar E(\bar{\mathbf{p}})$ comes from the harmonic frequency spectrum ${\omega_{n}}(\mathbf{\bar{p}})=n\bar{\omega}(\mathbf{\bar{p}})$
of a vibrating string with time periodicity $T_{t}(\bar{\mathbf{p}})$. This quantization is the field theory analogous  of  the semiclassical
quantization of a ``particle'' in a box, it also shares deep analogies
with the Matsubara theory, the Kaluza-Klein (KK) theory or with the Boltzmann kinetic theory. 
As discussed in the introduction the whole physical information of the system is
contained in a single space-time period $T^{\mu}$. Hence, the intrinsic space-time periodicity of a
free field describing the de Broglie ``periodic phenomenon'' can be represented as the field solution of a bosonic action in compact four dimensions
with PBCs \begin{equation}
\mathcal{S}=\oint_{0}^{T^{\mu}}d^{4}x\mathcal{L}(\partial_{\mu}\Phi,\Phi)\,.\label{free:act}\end{equation}
In this notation the circle in the volume integral symbols represents the PBCs. 
 It is important to note that the PBCs minimize the action at the boundaries.
 Therefore they have the
same formal validity of the usual (Synchronous) BCs of ordinary
field theory--- in particular along the time dimension. This is an essential feature  because it guarantees that
all the symmetries of the relativistic theory are preserved as in
ordinary field theory, \cite{Dolce:2009ce}.
In particular it guarantees that the theory is
Lorentz invariant. 

For instance we may consider a generic global Lorentz
transformation \begin{equation}
dx^{\mu}\rightarrow dx'^{\mu}=\Lambda_{\nu}^{\mu}~dx^{\nu}~,~~~~~~~~~~\bar{p}_{\mu}\rightarrow\bar{p}'_{\mu}=\Lambda_{\mu}^{\nu}~\bar{p}_{\nu}\,.\label{space:mom:Lorentz:tranf}\end{equation}
 The phase of the field  
is invariant under periodic translations $T^{\mu}$, moreover it is a Lorentz scalar (phase harmony \cite{1996FoPhL}). According to (1), in case of global transformations we have
$$ e^{-\frac{i}{\hbar}x^{\mu}\bar{p}_{\mu}}=e^{-\frac{i}{\hbar}(x^{\mu}+cT^{\mu})\bar{p}_{\mu}}~~~~~~~~~~ \rightarrow~~~~~~~~~~ e^{-\frac{i}{\hbar}(x'^{\mu}+cT'^{\mu})\bar{p}'_{\mu}}= e^{-\frac{i}{\hbar}x'^{\mu}\bar{p}'_{\mu}} \,.$$  
In this way we see that the space-time periodicity is a (space-like, tangent) contravariant
four-vector. It transforms under global Lorentz transformations as
 \begin{equation}
T^{\mu}\rightarrow T'^{\mu}=\Lambda_{\nu}^{\mu}~T^{\nu}\,.\label{period:Lorentz:tranf}\end{equation}
This can be also inferred by noticing
that after the transformation of variables (\ref{space:mom:Lorentz:tranf}),
the free action (\ref{free:act}) turns
out to have transformed integration region,  \begin{equation}
\mathcal{S}=\oint_{0}^{T'^{\mu}}d^{4}x'\mathcal{L}(\partial'_{\mu}\Phi,\Phi)\,.\label{Lorentz:trans:action}\end{equation}
 Therefore,  according to (1), in the new reference system the resulting four-periodicity
$T'^{\mu}$ of the field is actually given by (\ref{period:Lorentz:tranf}).
Action (\ref{Lorentz:trans:action}) describes the same ``periodic phenomenon'' in the new reference frame in which the resulting four-momentum $\bar{p}'_{\mu}$ is given by (\ref{space:mom:Lorentz:tranf}).

Since $T^\mu$ transforms as a tangent four vector \cite{Kenyon:1990fx}, the phase harmony (1) can be rewritten in the form $c \bar p_\mu = h / T^\mu$. In this way it is easy to see that  the underlying Minkowski metric induces the following  constraint on the dynamical global modulations of de Broglie four-periodicity \begin{equation}
\frac{1}{T_{\tau}^{2}}\equiv\frac{1}{T_{\mu}}\frac{1}{T^{\mu}}\,.\label{disp:rel:periods}\end{equation}
 In fact, considering the above notation, this is nothing
but the relativistic constraint  $$\bar{M}^{2}c^{2}=\bar{p}^{\mu}\bar{p}_{\mu}\,.\label{relat:disp:rel}$$

We will  denote by the bar sign the quantities related to the fundamental mode.   
    That is,
\begin{equation}
\bar \Phi(x) = \bar{\mathcal{N}} \bar \phi(x) =  \bar{\mathcal{N}}  e^{-\frac{i}{\hbar} \bar p_{\mu}  x^{\mu}}~. \label{field:exp:modes:cov} \end{equation}
For instance,
the Lagrangian of the fundamental mode $\bar{\Phi}(x)$ is 
\begin{equation}
\bar{\mathcal{S}} =\frac{1}{2}\int^{T^{\mu}}d^{4}x\left[\partial_{\mu}\bar{\Phi}^{*}(x)\partial^{\mu}\bar{\Phi}(x)-\bar M^{2}\bar{\Phi}^{*}(x)\bar{\Phi}(x)\right]\,.\label{action:fundmode:flat:statmass}\end{equation}
The circle in the integral symbol has been removed because this single mode solution can also be retrieved by assuming ordinary Stationary BCs at the boundary $T^{\mu}$.

Note that the fundamental mode $\bar \Phi(x)$ coincides with the mode of Klein-Gordon
field with energy $\bar{E}$ and mass $\bar{\mbox{M}}$ \footnote{From this correspondence to a Klein-Gordon mode it is possible to generalize the geometrodynamical analysis of the de Broglie periodicities that we will perform later  to ordinary field theory, \cite{Dolce:tune}.}.  This also means that the fundamental mode can be used to describe a classical particle. Thus, neglecting the higher harmonics of the field, the fundamental mode can be always quantized through second quantization obtaining ordinary quantum field theory. 
Nevertheless, without any explicit quantization, when all the possible energy eigenmodes allowed by the PBCs are considered, the periodic field turns out  to have 
 the same quantized energy spectrum as the corresponding ordinary second quantized
field --- after normal ordering. In fact, from (\ref{disp:rel:periods}) we find that the dispersion relation of the harmonic energy spectrum of (\ref{relat:disp:rel}) is \begin{equation}
\bar{E}_{n}(\mathbf{\bar{p}})=n \hbar \bar{\omega}(\mathbf{\bar{p}}) = n\sqrt{\bar{\mathbf{p}}^{2}c^{2}+\bar{M}^{2}c^{4}}\,.\label{Disp:rel:spectr}\end{equation}

Depending whether we want to make explicit the normalization factor of the energy eigenmodes or not,  the cyclic field solution can be described by the following notations,  \begin{eqnarray}
\Phi(x)  
= \sum_{n}  \Phi_{n}(x) =  
 \sum_{n}\mathcal{N}_{n}\alpha_{n}(\bar{\mathbf{p}})\phi_{n}(x) = \sum_{n}\mathcal{N}_{n}\alpha_{n}(\bar{\mathbf{p}})e^{-\frac{i}{\hbar} p_{{ n}} \cdot x}~. \label{field:exp:modes:norm} \end{eqnarray}
  In the free case, \emph{i.e.} persistent periodicity, the four-momentum spectrum is harmonic $p_{n \mu} = n \bar p_{\mu}$. That is, in the phase of the field, the PBCs yields   the Bohr-Sommerfeld condition for isochronic systems $\oint dx^{\mu} p_{n \mu}=p_{n \mu} T^{\mu} = n h$. The single quantum number $n$  is related to de fundamental periodicity $\mathbb S^{1}$ of the de Broglie clock. A complete description of the intrinsically periodic field solution however should involve a further expansion in spherical harmonics. For instance, in case of isotropic symmetry the field solution must be described in terms of two additional cyclic coordinates, \textit{i.e.} the angles $\theta \rightarrow \theta +  \pi$ and $\varphi \rightarrow \varphi + 2  \pi$. These further periodicities reproduce in the usual way the ordinary quantization of the angular momentum. As a consequence, the cyclic field solution has  two additional quantum numbers $\{m,l\}$ for a resulting topology $\mathbb S^{1}\otimes \mathbb S^{2}$. In this paper we will not consider any further the expansion in spherical harmonics or their deformations. {We also note that the natural generalization of this bosonic description to fermions is given by Schr\"odinger's \emph{Zitterbewegung} model,}  in which the spin and the magnetic momentum have a semi-classical interpretation in terms of de Broglie cyclic dynamics.

Finally, it is interesting to note 
that in the rest frame ($\mathbf{\bar{p}}\equiv0$) the quantized
energy spectrum (\ref{Disp:rel:spectr}) turns pout to be  \textit{dual} to the harmonic KK mass tower
$M_{n}=E_{n}(0)/c^{2}=n\bar{M}$. This mass spectrum is obtained by evaluating the harmonic Bohr-Sommerfeld condition in the rest frame: $M_{n} c \lambda_{s} = n h$. Indeed, for such a massive field,
the assumption of periodicity along the time dimension implies in the rest frame an intrinsic periodicity proper-time $\tau$. Its periodicity is
 \[
T_{\tau}=T_{t}(0)=\frac{h}{\bar{M}c^{2}}\,.\]
The invariant mass $\bar{M}$ is
 fixed geometrically by the reciprocal of the proper-time intrinsic
periodicity $T_{\tau}$ of the elementary field. In other words, by
imposing intrinsic time periodicity, the world-line parameter $s=c\tau$
turns out to be compact with PBCs\footnote{Since the theory can be even regarded as a simple type of string theory, it would be more appropriate to speak about strings rather than fields. In fact, the world-line parameter plays the role of  the compact
world-sheet parameter of ordinary string theory,  the PBCs  can be easily generalized to Neumann or Dirichlet BCs. 
}.  Its compactification length is $\lambda_{s}=cT_{\tau}$, \emph{i.e.}
 the Compton wave length of the field. The world-line parameter behaves similarly to the XD
of a KK field with zero 5D mass and compactification length is $\lambda_{s}=cT_{\tau}$. In order to bear in mind
these analogies with an XD field theory we say that the world-line
parameter play the role of a \emph{Virtual} XD (VXD). This aspect of the theory has been explored in the recent paper \cite{Dolce:AdSCFT}. It is interesting to note that, originally,
T. Kaluza introduced the XD formalism as a ``mathematical trick'' to unify gravity and electromagnetism
and not as a \emph{real} XD \cite{Kaluza:1921tu}, whereas the original O. Klein proposal was to use PBCs at the end of a compact XD (cyclic XD) in the attempt to interpret QM \cite{Klein:1926tv}.

\section{Quantum Behavior}\label{Quantum-Behavior}

Next we summarize  the correspondence between the  cyclic evolution of a periodic field with  the canonical formulation of QM as well
as with the Feynman Path Integral (FPI) formulation, as proven in \cite{Dolce:2009ce,Dolce:tune}. The evolution
along the compact time dimension is described by the so called bulk
equations of motion $(\partial_{t}^{2}+\omega_{n}^{2})\phi_{n}(x,t)=0$
--- for the sake of simplicity in this section we assume a single spatial
dimension $x$, avoiding the expansion in spherical harmonics. 
Thus the time evolution of the energy eigenmodes can be written as
first order differential equations $i\hbar\partial_{t}\phi_{n}({x},t)=E_{n}\phi_{n}({x},t)$.
 The periodic field (\ref{field:exp:modes:norm}) is a sum of on-shell
standing waves. Actually this harmonic classical system is the typical case where a Hilbert
space can be defined. In fact, the energy eigenmodes form a complete
set with respect to the inner product \begin{equation}
\left\langle \phi|\chi\right\rangle \equiv\int_{0}^{\lambda_{x}}\frac{{dx}}{{\lambda_{x}}}\phi^{*}(x)\chi(x)\,.\label{inner:prod}\end{equation}
 Therefore the energy eigenmodes  can be defined as Hilbert eigenstates
$\left\langle {x}|\phi_{n}\right\rangle \equiv{\phi_{n}({x})}/{\sqrt{\lambda_{x}}}$. As can be easily seen in the free case where the periodicity is persistent, this definition of the Hilbert space can be extended to an integral over an arbitrary large number $N_x$ of periods $\lambda_{x} \rightarrow V_x = N_x \lambda_x$. 
On this Hilbert space we can formally build a Hamiltonian operator $\mathcal{H}\left|\phi_{n}\right\rangle \equiv\hbar\omega_{n}\left|\phi_{n}\right\rangle $
and a momentum operator $\mathcal{P}\left|\phi_{n}\right\rangle \equiv-\hbar k_{n}\left|\phi_{n}\right\rangle $,
where $k_{n}=n\bar{k}=nh/\lambda_{x}$. Thus the time evolution of
a generic Hilbert state $|\phi(0)\rangle\equiv\sum_{n}a_{n}|\phi_{n}\rangle$, \emph{i.e.} of a generic cyclic field,
is actually described by the familiar Schr\"odinger equation \begin{equation}
i\hbar\partial_{t}|\phi(t)\rangle=\mathcal{H}|\phi(t)\rangle.\end{equation}
 Moreover the time evolution is given by the usual time evolution
operator $\mathcal{U}(t';t)=\exp[{-\frac{{i}}{\hbar}\mathcal{H}(t-t')}]$
which turns out to be a Marcovian (unitary) operator: $\mathcal{U}(t'';t')=\prod_{m=0}^{N-1}\mathcal{U}(t'+t_{m+1};t'+t_{m}-\epsilon)$
where $N\epsilon=t''-t'~$.
From the fact that the spatial coordinate is in this theory a cyclic
variable; by using the definition of the expectation value of an observable
$\partial_{x}F(x)$ between the generic initial and final states
$|\phi_{i}\rangle$ and $|\phi_{f}\rangle$ of this Hilbert space;
and integrating by parts (\ref{inner:prod}), we find \begin{equation}
\hbar \left\langle \phi_{f}|\partial_{x}F(x)|\phi_{i}\right\rangle = i \left\langle \phi_{f}|\mathcal{P}F(x)-F(x)\mathcal{P}|\phi_{i}\right\rangle \,.\end{equation}
 Assuming now that the observable is such that $F(x)=x$ \cite{Feynman:1942us}
we \emph{obtain the usual commutation relation of ordinary QM} $[x,\mathcal{P}]=i\hbar$ (without postulating it) --- more in general $[F(x),\mathcal{P}]=i\hbar \partial_x F(x)$.
With this result we have checked the correspondence with canonical
QM. 

To prove the correspondence with the ordinary
FPI formulation, it is sufficient to plug the completeness
relation of the energy eigenmodes in between the elementary time evolutions
of the Marcovian operator. With these elements at hand
and proceeding
in a complete standard way we find that the evolution of the cyclic fields turns out to be described by the usual FPI which in phase space is
 \begin{equation}
\mathcal{Z}=\lim_{N\rightarrow\infty}\int^{V_{x}}\left(\prod_{m=1}^{N-1}\frac{dx_{m}}{V_{x}}\right)\prod_{m=0}^{N-1}\left\langle \phi\right|e^{-\frac{i}{\hbar}(\mathcal{H}\Delta\epsilon_{m}-\mathcal{P}\Delta x_{m})}\left|\phi\right\rangle \,.\label{periodic:path.integr:Oper:Fey}\end{equation}

Proceeding in analogy with the ordinary derivation of the FPI in configuration space we also note  that the infinitesimal products in (\ref{periodic:path.integr:Oper:Fey}) can be generically written in terms of the action of the corresponding classical particle 
 \begin{equation}
{\mathcal{S}}_{cl}(t_{f},t_{i})\equiv\int_{t_{i}}^{t_{f}}dt {L}_{cl}  = \int_{t_{i}}^{t_{f}} dt \left(\mathcal{P}\dot{\mathrm{x}} - \mathcal{H} \right) \label{action:operator:generic}
\end{equation} 
Finally, the FPI (\ref{periodic:path.integr:Oper:Fey})
can be written in the familiar form ($\lim_{N\rightarrow\infty}\int^{V_{x}}\prod_{m=1}^{N-1}\frac{dx_{m}}{V_{x}} \equiv \int^{V_{\mathrm{x}}} \mathcal{D}\mathrm{x}$)
\begin{equation}
\mathcal{Z}=\int^{V_{\mathrm{x}}} \mathcal{D}\mathrm{x} e^{\frac{i}{\hbar}\mathcal{S}_{cl}(t_{f},t_{i})}\,.\label{eq:Feynman:Path:Integral}\end{equation}

\emph{This fundamental result has been obtained without any further assumption
than PBCs}. Here we have considered the case of a free ``periodic phenomenon'' being this case  sufficient to show the essential aspects of the correspondence. Nevertheless, as we will discuss later, the derivation of the FPI (\ref{eq:Feynman:Path:Integral}) can be generalized to interaction, see \cite{Dolce:tune}. 
The path integral description of the evolution of a ``periodic phenomenon''   has a simple classical interpretation, as can be shown in both a mathematical  and graphical way, \cite{Dolce:2009ce,Dolce:2009cev4}. In a cyclic geometry, such as that associated to a ``de Broglie periodic phenomenon'',
 there is an infinite set of possible classical paths with different
winding numbers that link every given initial and final configurations. 
Thus there are many possible classical evolutions of a field from
an initial configuration to a final configuration, which can interfere. The self-interference of a ``periodic phenomenon'' leads formally to  the ordinary FPI (\ref{eq:Feynman:Path:Integral}). It is important to note
 fundamental conceptual difference with respect to the usual Feynman
formulation: all these possible paths with different winding number are classical paths associated to the geometry $\mathbb S^{1}$. This means that in this path integral formulation it is not necessary to relax the classical variational principle in order to have path interference. 

The assumption of intrinsic periodicity enforces the wave-particle duality  of ordinary QM; it can be regarded as the literal realization of the de Broglie's original hypothesis of ``periodic phenomenon'' \cite{Broglie:1924}.
The non-quantum limit of the massive case, \emph{i.e.} the non-relativistic
single particle description, is obtained by putting the mass to infinity.
 As shown in \cite{Dolce:2009ce,Dolce:2009cev4}, in the 
classical limit only the first level of the spectrum must
be effectively considered. That is, in the  non-quantum limit the ``periodic phenomenon'' typically ``collapse'' to the ground state of four-momentum $\bar p_{\mu}$.  This yields a consistent interpretation of the double slit experiment \cite{Dolce:2009cev4}.   In fact, as can be seen by plotting the $|\Phi(x)|^{2}$ (factorizing the de Broglie internal clock), a massive cyclic field turns out to be localized within its Compton wavelength $\lambda_s$ along the corresponding classical path. In the classical-limit $\bar M \rightarrow \infty $, \emph{i.e.} $\lambda_s \rightarrow 0$, we get a Dirac delta distribution. Thus, at high frequencies, we typically pass from a wave to a particle description.  

    A massless field has infinite Compton wavelength and thus an infinite proper-time periodicity. Its quantum limit is at high frequency in which, in fact, the PBCs are important.    In this limit  we manifestly have a quantized energy spectrum. Thus, as in Planck's description of the black-body radiation we have no UV catastrophe.   In agreement with the experimental  observations, the corpuscular description (single photon) arises at high frequencies. The opposite limit is when time periodicity tends to infinity, so that  we have a continuous energy spectrum and a wave description (the thermal noise destroys the periodicity in a sort of ``decoherence''). 

The assumption of intrinsic periodicity implicitly contains
the Heisenberg uncertain relation of QM. Briefly, the phase of a `` de Broglie clock''  can not be determined and is  defined modulo factor $\pi n$ since only the square of the field has physical meaning, according to (\ref{inner:prod})--- this also bring the factor $1/2$ in the resulting uncertain relation. Thus to determine the energy
$\bar{E}(\mathbf{\bar{p}})=\hbar \bar{\omega}(\mathbf{\bar{p}})$ with good accuracy $\Delta\bar{E}(\mathbf{\bar{p}})$ we
must count a large number of cycles. That is to say we must observe the system
for a long time $\Delta t(\mathbf{\bar{p}})$ according to the relation $\Delta\bar{E}(\mathbf{\bar{p}})\Delta t(\mathbf{\bar{p}})\gtrsim\hbar/2$.    The simple mathematic demonstration of this relation is given in \cite{Dolce:2009ce}. 

The idea of using cyclic dynamics in the attempt to avoid hidden variables  is inspired to 't Hooft deteministic model \cite{'tHooft:2001ar}, where however the period $T_{t}$ is of order of
 of Planck time (and the Hamiltonian operator turns out to be not positive defined)  \cite{Nikolic:2006az}.   In our case, similarly to the KK theory where there are no tachyons, a cyclic field can have positive of negative frequency modes but the energy spectrum describes always positive energies and the Hamiltonian operator is positive defined.
The PBCs of the theory play the role of quantization condition; we have a fully relativistic generalization of the quantization of a particle in a box. Therefore we have the remarkable property that \textit{QM emerges without involving any hidden-variable}. Bell theorem can not be applied to our theory because the hypothesis of local and hidden variable are not satisfied (the assumption of intrisic periodicity can be regarded as an element of non locality in the theory).  As suggested by the formal correspondence with QM (in particular concerning the expectation value of an observable) described so far  the theory can in principle violates the Bell's inequality as QM.

\section{Geometrodynamics}

To introduce interactions we must consider that the four-periodicity
$T^{\mu}$ is fixed by the four-momentum $\bar{p}_{\mu}$,
according to the de Broglie-Planck relation (\ref{eq:PdB:relation}).
As already said, to an isolated elementary system (\textit{i.e.} free
field) can be associated a persistent fundamental four-momentum $\bar p_{\mu}$ and  four-periodicity $T^{\mu}$. On the other hand, an elementary
system under a generic interaction scheme, can be described in terms
of corresponding variations of four-momentum along its evolution with
respect to the free case. For instance the value of the four-momentum in a given interaction point $x=X$ can be written as \begin{equation}
\bar{p}_{\mu}\rightarrow\bar{p}'_{\mu}(X)=e_{\mu}^{a}(x)_{x=X}\bar{p}_{a}\,.\label{eq:deform:4mom:generic:int}\end{equation}
 In other words we describe interactions in terms of the so called
tetrad (or virebein) $e_{\mu}^{a}(x)$. Thus the interaction scheme
(\ref{eq:deform:4mom:generic:int}) turns out to be encoded in
the corresponding modulation  (variation) of the local (instantaneous) four-periodicity, which in the interaction point $x=X$ is \begin{equation}
T^{\mu}\rightarrow T'^{\mu}(X)= e_{a}^{\mu}(x)_{x=X}T^{a}\,.\label{eq:deform:4period:generic:int}\end{equation}
 Thus, in our formalism, the variation of periodicity occurring during interactions can be described
 as stretching of the compact dimensions of the theory. 
This suggests that the interaction scheme (\ref{eq:deform:4mom:generic:int}) can be equivalently
encoded into a corresponding locally curved space-time background
 \begin{equation}
\eta_{\mu\nu}\rightarrow g_{\mu\nu}(x)=e_{\;\mu}^{a}(x)e_{\nu}^{\; b}(x)\eta_{ab}\,.
\label{eq:deform:metric:generic:int}\end{equation}

This description can be checked by means of the following local transformation
of space-time variables \begin{equation}
x^{\mu}\rightarrow x'^{\mu}(x)=x^{a}\Lambda_{a}^{\;\mu}(x)\,, ~~ \text{where:} ~~~ 
e_{\;\mu}^{a}(x)=\left(\frac{\partial x^{a}}{\partial x'^{\;\mu}}\right)\,, ~~x^{a}\Lambda_{a}^{\;\mu}(x)\simeq\int^{x^{a}}dx^{a}e_{a}^{\;\mu}(x)\,. \label{eq:deform:mesure:generic:int}\end{equation}
 We are working in the approximation $e_{\;\mu}^{a}(x')\simeq e_{\;\mu}^{a}(x)$ (for the sake of simplicity we neglect Christoffel symbols). 
In the last relation we have used the fact that  $x^{a} \Lambda_{a}^{\;\mu}(x)$ is the primitive
of the tetrad $e^{\;\mu}_{a}(x)$  (omitting
prime indexes in the integrand). Indeed, by using   (\ref{eq:deform:mesure:generic:int})    as a substitution of variables in 
 the compact 4D action  (\ref{free:act}), we find that the interaction scheme (\ref{eq:deform:4mom:generic:int}) is
 described by the following local action  in deformed compact 4D 
\begin{equation}
\mathcal{S}\simeq\oint^{T^{a}\Lambda_{a}^{\mu}|_{X}(T)}d^{4}x\sqrt{-g(x)}\mathcal{L}(e_{a}^{\;\mu}(x)\partial_{\mu}\Phi'(x),\Phi'(x))\,.\label{eq:defom:action:generic:int}\end{equation}
The transformation (\ref{eq:deform:mesure:generic:int}) relates locally the inertial frame $x \in  S$ of the free cyclic field solution $\Phi(x)$   to the generic frame $x' \in  S'$ associated with the interacting cyclic field solution $\Phi'(x)$. Its Jacobian is  $\sqrt{-g(x)}=\det[e_{\;\mu}^{a}(x)]$. The Latin letters describe the free field 
 in an inertial frame $ S$ while
the Greek letters refer to the 
 locally accelerated frame $ S'$ of the interacting field $\Phi'$ \cite{Birrell:1982ix}. 

It is important to point out that (\ref{eq:deform:mesure:generic:int}) induces 
the local deformation (or stretching) of the boundary 
\begin{equation}
 T^{a}\Lambda_{a}^{\;\mu}|_{X}(T) \simeq  \int^{X^{a}+T^{a}}_{X^{a}} dx^{a} e_{a}^{\;\mu}(x)\,.\label{eq:deform:4T:generic:int}
 \end{equation}
 This actually is the local deformation of the boundary associated with the modulation of local periodicity $T'^{\mu}(x)$, (\ref{eq:deform:4period:generic:int}), \emph{i.e.} with the interaction scheme (\ref{eq:deform:4mom:generic:int}).
With this local transformations of variable we see that  the boundary (\ref{eq:deform:4T:generic:int}) transforms  as a ordinary four-vector ($\propto x^{\mu}$) whereas the related periodicity (\ref{eq:deform:4period:generic:int}) transforms as a tangent four vector ($\propto dx^{\mu}$), \cite{Kenyon:1990fx}.    

This geometrodynamical approach to interactions is interesting because
it actually mimics  the usual geometrodynamical approach
of GR. Actually gravitational interaction can be described in terms of variation of periodicity of reference clocks. In fact, considering only the fundamental mode (\emph{i.e.} neglecting quantum corrections), if we suppose a weak Newtonian potential $V(\mathbf{x})=-{GM_{\odot}}/{|\mathbf{x}|}\ll c^{2}$,
we find that the energy of a de Broglie ``internal clock'' on a gravitational well varies (with respect
to the free case) as $\bar{E}\rightarrow\bar{E}'\sim\left(1+{GM_{\odot}}/{|\mathbf{x}| c^{2}}\right)\bar{E}$.
According to (\ref{eq:deform:4period:generic:int}) or (\ref{eq:PdB:relation}),
this means that the de Broglie clocks in a gravitational well are
slower with respect to the free clocks $T_{t}\rightarrow T_{t}'\sim\left(1-{GM_{\odot}}/{|\mathbf{x}| c^{2}}\right)T_{t}$.
Thus we have a gravitational redshift $\bar{\omega}\rightarrow\bar{\omega}'\sim\left(1+{GM_{\odot}}/{|\mathbf{x}| c^{2}}\right)\bar{\omega}$.
With this schematization of interactions we have retrieved two important
predictions of GR.
Besides the time periodicity we must also consider the analogous variation of spatial
momentum and the corresponding contraction of spatial wavelength of the particles. 
The weak Newtonian interaction turns out to be encoded in the usual linearized Schwarzschild metric \cite{Ohanian:1995uu}, in agreement with our description (\ref{eq:deform:metric:generic:int}).  
 We have found that the geometrodynamical approach to interactions of a ``periodic phenomenon''
actually mimics linearized gravity.
Here we only mention that, as well known, see for instance \cite{Ohanian:1995uu}, it is possible
to retrieve ordinary GR from such a linear formulation by including self-interactions. Furthermore, it is important to mention that it is not uniquely defined  ``what is fixed at the boundary of the action principle of GR'' \cite{springerlink:10.1007/BF01889475}. SR and GR fix the differential framework of the 4D theory without giving any particular prescription about the BCs. The only requirement for the BCs is to minimize the relativistic action at the boundary. For this aspect both  PBCs have the same formal validity and consistence with relativity as the  Stationary BCs of ordinary QFT.

We conclude that an elementary system under the interaction scheme (\ref{eq:deform:4mom:generic:int})
is described by the modulated solution of the bulk equations of motion  on the deformed compact background (\ref{eq:deform:metric:generic:int})
and intrinsic periodicity (\ref{eq:deform:4period:generic:int}). 
Intuitively we can mention that, \cite{Dolce:tune},  this modulation of periodicity for single energy eigenmodes  can be described by the formalism of modulated signals, \emph{e.g.}, the frequency modulation  in the phase factor of the free fundamental mode is: $-\frac{i}{\hbar}x^{\mu} \bar{p}_{\mu} \rightarrow -\frac{i}{\hbar}\int^{x'} d x^{\mu}  p'_\mu(x)$. As discussed in detail in \cite{Dolce:tune}, in doing this we are assuming that the normalization factor is invariant. Similarly to the Bohr-Sommerfeld condition we find that, in a given interaction point $x=X$, the quantization condition of the the modulated solution  coming from the transformed PBCs is given by $c T^{\mu} p_{n \mu}= h n \rightarrow  \oint_X d x^\mu  \bar p'_{n \mu}(x) = n h$.  
 From this, it can be shown that the space-time evolution of an  interacting ``periodic phenomenon'' is Markovian and  the Hamiltonian and momentum operator transform as $\mathcal{P}_{\mu}\rightarrow\mathcal{P}'_{\mu}(X)=e_{\mu}^{a}(x)_{x=X}\mathcal{P}_{a}$, where the four-momentum operator is  $\mathcal{P}_{\mu} \equiv \{ \mathcal H/c, - \mathcal P_{i}\}$.  This means that  the FPI of the interacting system is  obtained through the formal substitution 
$\mathcal{S}(t_{f},t_{i}) \rightarrow  \mathcal{S}'(t_{f},t_{i})$ in  (\ref{eq:Feynman:Path:Integral}). From the definition of $\mathcal{P}'_{\mu}(X)$ it follows that  $\mathcal{S}'_{cl}(t_{b},t_{a})$ is formally the action of the corresponding interacting classical particle (in terms of operators). The integral $\int^{V_{x}} \mathcal{D} {x} $ takes into account that in every point of the evolution of an interacting cyclic field a different inner product (\ref{inner:prod}), and a thus different Hilbert space, is defined (the spatial periodicity varies locally). We have also assumed that  the integration volume $V_{x}$ is bigger than the interaction region $\mathcal{I}$ or infinite, \emph{i.e.}  a large or infinite number of periods $N$ so that it can be considered overall not affected by local deformations.

\section{Generalization to gauge interactions}

This section is a short introduction to the results of the recent study about the space-time geometrodynamics of gauge interactions \cite{Dolce:tune}.  
For the sake of simplicity the geometrodynamics investigated are approximated to local isometries, $\sqrt{-g'} \simeq 1$.  The equation of motions  (KG equation in ordinary field theory) are left invariant under local transformations of flat space-time.
Thus, in ordinary field theory local  isometries   have no effect on the field solution.  In fact, the KG field used for practical computations  is the most generic solution of the KG equation; BCs have a marginal role in ordinary field theory computations.  However  local transformations of flat metric affect the boundary of the theory. For instance we can have local rotations of the boundary of the theory described by (\ref{eq:deform:4T:generic:int}), whereas the structure of the Lagrangian is left invariant, see (\ref{eq:defom:action:generic:int}).
Actually, one of the interesting characteristics of the formalism of compact space-time dimensions is that it allow us to control the variations of the boundary (\ref{eq:deform:4T:generic:int}) and the consequent variations of field solution associated with transformations of the reference frame (\ref{eq:deform:mesure:generic:int}). 
That is,  the resulting local rotation of BCs of the field implies a local variation of the solution in field theory in compact space-time dimensions.  In \cite{Dolce:tune} we have shown that such local variations of field solution induced by local transformations of reference frame can be identified with the internal transformations of ordinary gauge theory. In particular the interaction scheme (\ref{eq:deform:4mom:generic:int}) corresponding to these local isometries (the  case of local  conformal invariance  has been partially investigated through the dualism with XD theories in \cite{Dolce:AdSCFT})
    can be formally written as the ordinary minimal substitution.      
In this way it is possible to see that the resulting gauge field encodes the local modulations of periodicity associated to the local transformation of variables. 
  We have also shown that fields with different periodicities are allowed  in an action with persistent boundary as long as they appear in gauge invariant terms. In fact gauge transformations tune the periodicity of the different fields of the theory, so that the action can be minimized at the common boundary.  This implies that only particular types of  local isometries, which we call  \emph{polarized}, are consistent with  the variational principle. These polarized  local isometries are actually those reproducing Maxwell dynamics for the gauge field.  Indeed, the dynamics associated with these particular local  isometries reveal a geometric space-time nature of gauge interaction \cite{Dolce:tune}.   
 This  can be regarded as in the spirit  of  Weyl's and Kaluza's original proposals.

The PBCs at the geometrodynamical boundary of field theory in compact space-time dimensions represent a semi-classical quantization condition, relativistic generalization of the quantization of a particle in a box \cite{Dolce:2009ce,Dolce:AdSCFT}.
  The demonstrations of sec.(\ref{Quantum-Behavior}) can be generalized to gauge interaction. When the PBCs are explicitly imposed at the geometrodynamical boundary of the theory,
 we  find that the modulation of four-periodicity of all the energy eigenmodes constituting an interacting  cyclic field  turns out to define different local Hilbert spaces and formally described by the ordinary Scattering Matrix of QM. Similarly, the evolution associated with such a locally modulated  ``periodic phenomenon'' turns out to formally correspond  with the ordinary FPI of scalar QED.    

\section{Conclusions}

In this paper we have presented a semi-classical interpretation of quantum field theory based on the de Broglie's postulate of ``the existence of periodic phenomena allied with each parcel of energy'' \cite{Broglie:1924}.   Such an assumption of intrinsic ``periodic phenomenon'' is implicitly at the base of the ordinary undulatory formulation of relativistic QM and has been indirectly observed in a recent experiment \cite{2008FoPh...38..659C}.  Here we have formalized intrinsic periodicity by using the formalism of field theory in compact space-time dimensions and PBCs \cite{Dolce:2009ce}.  The main requirement for the BCs is to fulfill the variational principle. Actually, PBCs minimize the relativistic bosonic action at the boundaries of compact space-time dimensions. Therefore, the  assumption of intrinsic periodicity is fully consistent with a relativistic bosonic theory.  Relativistic causality and time ordering  are guaranteed by the fact that the space-time periodicity $T^\mu$ of a de Broglie ``periodic phenomenon''  and the four-momentum $\bar p_\mu$ of the corresponding relativistic elementary system are dynamically related through the de Broglie-Planck relation (1).  Indeed, the local, retarded variations of four-momentum characterizing relativistic interactions  are equivalently described by corresponding  local, retarded modulations of the de Broglie space-time periodicity.  In the theory this is represented by the fact that the boundary and the metric vary in a relativistic way. This also means that in our formalism interactions can be described in terms of space-time geometrodynamics.  As shown in \cite{Dolce:tune}, this formalism reveals the geometrodynamical nature of gauge interactions,  analogous to that of gravitational interaction.  The gauge field turns out to encode the modulations of periodicity of particular interacting ``periodic phenomena''. Moreover, under the assumption of intrinsic periodicity an elementary particle can be regarded as a vibrating string, leading to the full relativistic generalization of the semi-classical quantization of the particle in a box. As a result, without introducing any hidden variable, such a classical theory has remarkable formal correspondences to the fundamental aspects of ordinary relativistic QM. ``\emph{This hypothesis [of periodic phenomenon] is at the base of our theory: it is worth as much, like all hypotheses, as can be deduced from its consequences}'' \cite{Broglie:1924}. We conclude that, after nearly 90 years, de Broglie's ideas can play a renewed role to address fundamental, conceptual and computational open questions of modern physics.     

\section*{References}
\providecommand{\newblock}{}

\end{document}